\documentclass[10pt,twocolumn]{article}
\pdfoutput=1
\newif\ifOverleaf
\Overleaffalse
\usepackage{times}
\usepackage{fullpage}
\usepackage{algorithm}
\usepackage{algpseudocode}
\usepackage{amsmath}
\usepackage{amsthm}
\usepackage{amsfonts}
\usepackage{amssymb}
\usepackage{array}
\usepackage{appendix}
\usepackage{booktabs}
\usepackage{boxedminipage}
\usepackage{calrsfs}
\usepackage{caption}
\usepackage{color}
\usepackage{enumitem}
\usepackage{fancyhdr}
\usepackage{float}
\usepackage[T1]{fontenc}
\usepackage{graphicx}
\PassOptionsToPackage{hyphens}{url}
\usepackage{hyperref}
\hypersetup{
    colorlinks=false,
    pdfborder={0 0 0},
}
\usepackage[capitalize]{cleveref}
\usepackage[utf8]{inputenc} 
\usepackage{listings}
\usepackage[english]{babel}
\usepackage{lmodern}
\usepackage[activate={true,nocompatibility},final,tracking=true,kerning=true,spacing=true,factor=1100,stretch=10,shrink=10]{microtype} 
\usepackage{multirow}
\usepackage{nicefrac}
\usepackage{pifont}
\usepackage[section]{placeins}
\usepackage{sidecap}
\usepackage{siunitx}
\usepackage{subcaption}
\usepackage{verbatim}
\usepackage[usenames,dvipsnames]{xcolor}
\usepackage{xspace}
\usepackage[ff,sets,adversary,keys,primitives,operators,lambda,logic,notions]{style/cryptocode}

\lstdefinelanguage{javascript}{
  keywords={typeof, new, true, false, catch, function, return, null, catch, switch, var, if, in, while, do, else, case, break},
  keywordstyle=\color{blue}\bfseries,
  ndkeywords={class, export, boolean, throw, implements, import, this},
  ndkeywordstyle=\color{black}\bfseries,
  identifierstyle=\color{black},
  sensitive=false,
  comment=[l]{//},
  morecomment=[s]{/*}{*/},
  commentstyle=\color{purple}\ttfamily,
  stringstyle=\color{red}\ttfamily,
  morestring=[b]',
  morestring=[b]"
}

\newcommand{\squishlist}{\begin{itemize}[itemsep=0pt,parsep=0pt,topsep=0pt,partopsep=0pt,leftmargin=1em,labelwidth=1em,labelsep=0.5em]}
\newcommand{\squishlistend}{\end{itemize}}
\newcommand{\squishend}{\end{itemize}}
\newcommand{\squishenum}{\begin{enumerate}[itemsep=0.5pt,parsep=0pt,topsep=0pt,partopsep=0pt,leftmargin=1.5em,labelwidth=1em,labelsep=0.5em]{}}
\newcommand{\squishenumend}{\end{enumerate}}

\newcommand\myurl[2]{\url{#1}}

\renewcommand{\floatpagefraction}{0.95}

\newcommand{\captionfonts}{\small}
\makeatletter  
\long\def\@makecaption#1#2{%
  \vskip 0.1in
  \sbox\@tempboxa{{\captionfonts #1: #2}}%
  \ifdim \wd\@tempboxa >\hsize
    {\captionfonts #1: #2\par}
  \else
    \hbox to\hsize{\hfil\box\@tempboxa\hfil}%
  \fi
  \vskip 0in}
\makeatother   

\newtheorem{theorem}{Theorem}

\newcommand{\Goober}{Leash\xspace}
\newcommand{\goober}{leash\xspace}
\newcommand{\Goobers}{Leashes\xspace}
\newcommand{\goobers}{leashes\xspace}
\newcommand{\gooberfunc}{\ensuremath{\textsf{leash}}\xspace}
\newcommand{\goobered}{leashed\xspace}
\newcommand{\goobering}{leashing\xspace}
\newcommand{\Goobered}{Leashed\xspace}
\newcommand{\Goobering}{Leashing\xspace}

\def\systemnameRaw{Short \Goobers}
\def\systemname{\systemnameRaw\xspace}

\newcommand{\Victim}{Alice\xspace}
\newcommand{\VictimLetter}{\ensuremath{\pazocal{A}}\xspace}
\newcommand{\VictimLong}{Alice Van Winkle\xspace}
\newcommand{\VictimTP}{she\xspace} 
\newcommand{\VictimTPP}{her\xspace} 
\newcommand{\VictimTPO}{her\xspace} 
\newcommand{\VictimTPCap}{She\xspace} 

\newcommand{\Extractor}{Cobb\xspace}
\newcommand{\ExtractorLetter}{\ensuremath{\pazocal{C}}\xspace}
\newcommand{\ExtractorLong}{Dom Cobb\xspace}
\newcommand{\ExtractorTP}{he\xspace}

\newcommand{\ExtractorTPO}{him\xspace}

\newcommand{\ExtractorTPCap}{He\xspace}

\newcommand{\Set}[1]{\ensuremath{\textsf{#1}}}
\newcommand{\DbState}[1]{\ensuremath{\text{\textbf{#1}}}}
\newcommand{\TxSeq}[1]{\ensuremath{\text{\textbf{#1}}}}
\newcommand{\WAmt}[1]{\ensuremath{\pazocal{#1}}}
\newcommand{\TxSet}[1]{\ensuremath{\text{\textsf{#1}}}}
\newcommand{\UintEvm}{\ensuremath{\mathbb{Z}_{2^{256}}}}
\newcommand{\Txn}{\Set{Txn}}

\newcommand{\blockhash}{\texttt{BLOCKHASH}\xspace}
\newcommand{\blockhashex}{\texttt{BLOCKHASH\_FD}\xspace}
\newcommand{\chainid}{\texttt{CHAINID}\xspace}

\newcommand{\depth}{\operatorname{\textsf{depth}}}
\newcommand{\distance}{\operatorname{\textsf{dist}}}
\newcommand{\isAncestorOf}{\operatorname{\textsf{isAncestorOf}}}
\newcommand{\parent}{\ensuremath{p}\xspace}
\newcommand{\up}{\textsf{up}}

\usepackage{svg}

\newcommand\blfootnote[1]{%
  \begingroup
  \renewcommand\thefootnote{}\footnote{#1}%
  \addtocounter{footnote}{-1}%
  \endgroup
}

\floatstyle{boxed}
\floatname{Workflow}{Listing}  
\newfloat{Workflow}{htb}{lop}

\newcommand{\mytilde}{\raise.17ex\hbox{$\scriptstyle\mathtt{\sim}$}}

\makeatletter
\@ifundefined{showcommenttrue}{\newif\ifshowcomment\showcommenttrue}{}%
\makeatother%
\newcommand{\NewReviewer}[2]{%
\ifshowcomment%
 \expandafter\newcommand\csname #1\endcsname[1]{\textsf{\color{#2}{[#1: {##1}]}}}%
\else
 \expandafter\newcommand\csname #1\endcsname[1]{}%
\fi%
}

\definecolor{bsyColor}{rgb}{0.50, 0.1, 0.8} 
\NewReviewer{bsy}{bsyColor}
\NewReviewer{dawn}{blue}
\NewReviewer{peter}{cyan}

\ifshowcomment
\newcommand{\todo}[1]{\textsf{\color{red}{[{TODO: #1}]}}}

\else
\newcommand{\todo}[1]{}

\fi

\showcommentfalse

\usepackage{balance}

\usepackage{txfonts}
\usepackage{calrsfs}  
\DeclareMathAlphabet{\pazocal}{OMS}{zpml}{m}{n}

\ifOverleaf 
\newcommand\githash{OVERLEAF-DRAFT-NO-GIT-HASH\xspace}

\newcommand\gitdate{\today\xspace}
\else
\newcommand\githash{692d\-3660\-6a22\-19e9\-d6fc\-790b\-2dc0\-07c2\-9fab\-190f\xspace}
\newcommand\gitdate{Jun 2, 2022\xspace}

\fi

\date{\gitdate}

\begin{document}

\title{Keep Your Transactions On \systemname\par
  {\Large or}\par
  {\Large Anchoring For Stability In A Multiverse of Block \emph{Tree} Madness}}

\author{
{\rm
  Bennet Yee
} \\
{\small
  Oasis Labs
}
} 

\maketitle


\ifshowcomment%
\section*{Reviewer Instructions}
\textbf{%
  {\color{red}DO NOT USE OVERLEAF COMMENTS.} %
  Use \LaTeX-style comments---use the $\backslash$\texttt{NewReviewer}
  macro in \texttt{reviewers.tex}.  Having comments and suggested
  changes in the text makes it possible to synchronize with
  \texttt{github} and edit outside the browser, and having the changes
  in the \texttt{git} history at \texttt{github} rather than squashed
  at overleaf makes it possible to revert previously accepted changes,
  e.g., when the change is no longer appropriate/applicable as a
  result of a later suggestion.}

\textbf{%
  Please pull from \texttt{github.com} to ensure that the overleaf
  copy is fresh (\texttt{Menu > github > pull github changes into
    overleaf}) before making comments.  It would also help if you put
  your comment on its own line, end with a \texttt{\%} to prevent text
  reflows from combining it with the next line to reduce the
  likelihood of merge conflicts, and pushed your changes to
  \texttt{github} after you make them (\texttt{Menu > github > push
    overleaf changes to github}).}
\fi

\section*{Abstract}

The adversary's goal in mounting Long Range Attacks (LRAs) is to fool potential victims into using and relying on a side chain, i.e., a false, alternate history of transactions, and into proposing transactions that end up harming themselves or others.
Previous research work on LRAs on blockchain systems have used, at a high level, one of two approaches.
They either try to (1) prevent the creation of a bogus side chain or (2) make it possible to distinguish such a side chain from the main consensus chain.

In this paper, we take a different approach.
We start with the indistinguishability of side chains from the consensus chain---for the eclipsed victim---as a given and assume the potential victim \emph{will} be fooled.
Instead, we protect the victim via harm reduction applying ``short \goobers'' to transactions.
The \goobers prevent transactions from being used in the wrong context.

The primary contribution of this paper is the design and analysis of \goobers.
A secondary contribution is the careful explication of the LRA threat model in the context of BAR fault tolerance, and using it to analyze related work to identify their limitations.

\section{Introduction}
\label{sec:Introduction}

\blfootnote{This paper was built from \LaTeX\xspace source at git hash \texttt{\githash}.}

Smart contract blockchain systems aspire to be reliable computing---and economic---infrastructure that can be trusted to operate for a long time.
One concern expressed in the literature is the class of so-called Long Range Attacks (LRAs)~\cite{ButerinLongRangeAdaptiveProofOfWork,8525396,Deirmentzoglou2019ASO}.

LRAs are a form of ``alternate history attacks.''
The goal of the adversary is to convince a victim to act based on transactions logged to a side chain---a history of an alternative, fictional universe---rather than on those logged on the global consensus chain.

All LRAs are easily prevented on BFT-based, instant-finality, Proof-of-Stake (PoS) blockchains like the Oasis network~\cite{OasisBlockchainPlatformWhitepaper}.
As long as the users are following the blockchain's evolution as a ``light client'', they will know the correct global consensus state.
A light client continuously monitors block production and tracks the consensus protocol, verifying the metadata that pertains to consensus: checking the basic block structure of new blocks, including changes in consensus committee membership (who can sign new blocks), without validating the ``payload'' portion.
Here by ``payload'' data we mean the portion that contains the details about the evolution of the smart contract virtual machine state---transaction parameters, state root hashes, etc.
Light clients do not maintain a copy of smart contract persistent storage state---that would be expensive---and thus cannot verify the state root in the consensus blocks actually represents the result of transaction execution from the previous state; all they can verify is that the validation committee reached consensus on the blocks and that changes to the validation committee are themselves consensus driven.

While being able to act as light client is sufficient for preventing LRAs, it is not necessary.
Even if an intended victim cannot themselves be a light client, as long as they can trust $m$ out of $n$ others who \emph{are} light clients and can query them for what they believe to be the current blockchain state, that suffices to learn the true current consensus state, given proper selections of the $m$ and $n$ security (and availability) parameters.

When LRA targets cannot communicate to directly or indirectly learn about the blockchain's evolution, however, they are said to be ``eclipsed''; protecting eclipsed targets seems particularly difficult.
Other research in the LRA literature have attempted to prevent the creation of side chain or somehow make them detectable, and these approaches have serious practical limitations.
Instead, this paper takes the approach of minimizing the harm that might occur when a LRA victim is fooled into signing new transaction proposals.
We call our design a ``short \goober,'' since it essentially prevents a transaction from doing anything outside a short ``range'' of an identified blockchain state.

Next, we first more precisely state the threat model.
In Section~\ref{sec:ShortRangeLeashes}, we explain short \goobers in detail and show how they can be used to address posterior corruption / LRAs.
Following this, in Section~\ref{sec:RelatedWork} we discuss other approaches to address LRAs and other related work.
We provide some concluding remarks in Section~\ref{sec:Conclusion}. 

\section{Threat Model}
\label{sec:ThreatModel}

In this paper, we focus on the situations where gaining trustworthy information about the blockchain is impossible, i.e., where a victim is eclipsed so the adversary has full control over what messages can get through~\cite{cryptoeprint:2015:263}.
We first describe the LRA posterior corruption setting.
Next, we specify the security model, detailing what assumptions are made and clearly stating what the adversary is allowed (and not allowed) to do.

\subsection{Terminology / Setting}

Informally, the blockchain participant victim, whom we'll call \VictimLong (\VictimLetter), has been inactive or asleep and does not know what is the current state of the blockchain---or necessarily even know how much time has elapsed.
Furthermore, the adversary, \ExtractorLong (\ExtractorLetter), has the ability to control \Victim's (network) communications, deciding on which messages are blocked and which are allowed through.

Additionally, because the PoS blockchain operates for a long time, there will be past participants who have exited the ecosystem.
They have nothing at stake, and would not put any effort into protecting their old keys---and may even sell them.
Thus, keys belonging to some participants, including a super-threshold number of validators who served together in a past committee, will be cheaply available to the adversary.
This means that \Extractor is able to sign and forge blocks to create a side chain that forks from the blocks added when that validation committee was active.
\Victim, having recently awaken and also eclipsed, had not been keeping track of consensus state, would find this side-chain indistinguishable from the consensus chain and would happily accept it.
These kind of attack is also referred to as ``posterior corruption'', where history is (apparently) changed after the fact.

While \Victim remembers what \VictimTP knew before going to sleep and may have trusted others then, there is no other parties whom \Victim trusts \emph{after waking}---\VictimTP is only willing to trust the protocol design and implementation, since any other parties may have exited and sold their signature keys since \VictimTP fell asleep.
Note that there is no requirement that \Victim only wakes up once or that \Extractor only forge blocks once; as with cryptanalytic attacks, \ExtractorTP can be an adaptive adversary, e.g., creating bogus history in phases in attempting to get \VictimTPP to take more risks each time.

How can the newly awaken \Victim learn the true current state of the blockchain based only on what messages \Extractor allows through and \VictimTPP knowledge of the state of the blockchain at the time that \VictimTP went to sleep?

This is quite daunting.

While we are mainly concerned with Proof-of-Stake (PoS) chains, this is a problem for Proof-of-Work (PoW) blockchains too.
Here, creating a sidechain is ``easy'': \Extractor just needs hashing power.
Normally, the security assumption for PoW blockchains is that adversaries cannot acquire more than 50\% of global hashing power, and all participants rely on the longest chain rule to determine what is consensus transaction history.
However, \Victim being recently awakened and fully eclipsed means that \Extractor can prevent \VictimTPP from seeing any blocks from the consensus or main chain added after \VictimTP went to sleep.
And not knowing the elapsed time (with estimated block production rate), \VictimTP has no basis to guess what block height might be considered normal and what might be suspicious.
The notion of ``longest'' is a property that requires \emph{global knowledge}, and being eclipsed prevent \VictimTPP from acquiring information that might distinguish a shorter side chain from the main chain.

Note that \Extractor can create an alternate history diverting from the true consensus log at \emph{any} point after the last block \Victim saw before sleeping.
It is easier for \ExtractorTPO to do so as far back in time as possible for common PoW blockchains---right at the genesis block, if \VictimTP only knows about the system design, has the software, and the system configuration parameters such as the genesis block, but has not tracked the chain history at all.
This is because typical configurations of the hash parameters make it easier to create blocks when the chain height is low, and block creation becomes harder as the chain gets longer.

On PoS blockchains, building an alternate history that an eclipsed individual would accept require getting a quorum of validator signatures on the alternate history blocks.
The long-range aspect can greatly reduce the adversarial cost for mounting the attack: entities who served as validators but who are no longer participating in the ecosystem (have sold their tokens) will have nothing at stake, and under the BAR (Byzantine, Altruistic, Rational) model~\cite{aiyer2005bar} might even be willing to sell their old signing keys.
The more time has passed, the greater number of participants' keys will be available due to exiting the ecosystem or to key rotation.

\subsection{Adversary Capabilities}

For LRA posterior corruption, we assume that most cryptographic keys should be considered exposed.
We assume, however, that those keys belonging to \Victim, the intended victim, and the honest subset of the current/recent validation committees are not compromised.
However, recall that \VictimTP does not know the validation committee membership, so assumptions about the current (or recent) validation committee members' honesty does not help \VictimTPP.

This is the \emph{weakest possible set of assumptions}!
\Victim's key must be secure, since otherwise \Extractor can arbitrarily sign bogus transactions in \VictimTPP name.
If a super-threshold number of the current validation committee lost control of their keys, then the blockchain itself can be arbitrarily corrupted.
This similarly applies to recent validation committee members, exactly how many validation committee elections are covered by ``recent'' above is a security parameter.
Obviously, if this number is too low and users tolerate a level of communication delay that includes validation committee changes, then an adversary who has control over the communication network could fork the blockchain network: they could block legitimate messages from leaving the validation committee until enough elections occur, then use now non-recent validation committee member keys to generate bogus (but correctly signed) messages to create the side chain.

Note that we continue to assume that the cryptographic signature scheme to be secure, and that the cryptographic hash function used to link blocks together remain collision resistant.
\Extractor, the adversary, is allowed full access to cryptographic keys belonging to everyone else.

\subsection{Security Assumptions}

As stated above, we assume that the PoS blockchain use secure cryptographic schemes for signatures and for hashing.
We assume that even though \Victim may have slept decades, no critical cryptanalytic attacks have been discovered that would non-negligibly weaken security.

We do \emph{not} assume a secure source of time.
While Network Time Security~\cite{RFC8915} can be used to learn about the current time, it is not yet widely used and it can be completely blocked by the eclipsing adversary.
Worse, because of the potential passage of time, the NTS server's TLS private key---corresponding to a public key in an expired server certificate---may be compromisable in much the same way that validator keys are compromisable.

While \Victim can check signatures and authenticated data structures, \VictimTP does not know the identity of the current validator set.

\section{Short \Goobers}
\label{sec:ShortRangeLeashes}

The key idea in the short \goobers approach is to re-frame the LRA problem, so that instead of trying to detect side chains or prevent their creation, we try to nullify the potential damage that might result.
At the highest level, the adversary is attempting to defraud their potential victim(s) by presenting lies to them (on its sidechain) and convincing them to submit transactions based on those lies.
The short \goobers approach is orthogonal to side-chain detection / prevention; instead, it \emph{reduces the dynamic scope} of the user's transactions so they are not universally applicable.
Next, we'll present the notation that we'll use to discuss short \goobers and compare with other approaches.

\subsection{Notation and Terminology}

Our formalism models an Ethereum-style blockchain virtual machine, but can be adjusted for other blockchains.

As before, \Victim (\VictimLetter) is the potential victim, and \Extractor (\ExtractorLetter) is the adversary mounting the LRA to fool \VictimLetter into signing an inappropriate ``transaction proposal''.
We distinguish a proposed transaction from one that is committed to the blockchain, since users of blockchain systems often sign a second proposal with the same nonce but with different parameters (esp gas price) to try to ``cancel'' the earlier submitted-but-not-committed transaction still pending in the mempool.

The ``database state'' here refers to the persistent storage associated with a blockchain system.
More abstractly, it is the virtual machine state that results from executing all the finalized transactions, starting with the genesis state.
The state is implicitly determined by the logged transactions and the virtual machine definition, so the database contents can always be recomputed; for efficiency, it is explicitly represented.

More concretely, \Set{DbState} is a mapping $\Set{Account}\rightarrow\Set{AccountState}$, where \Set{Account} is a 256-bit value that is either a public key hash or a contract account number, derived from the contract creator's account and nonce, and where \Set{AccountState} is a tuple $\left(\Set{Nonce}, \Set{Balance}, ..., \Set{ContractState}\right)$.
Those accounts associated with public keys are user or Externally Owned Accounts (EOAs) that can initiate top-level transactions; the rest are associated with code---a smart contract---which executes invoked transactions and can in turn make subtransaction invocations.
Here \Set{Nonce} and \Set{Balance} are 256-bit values, etc, and \Set{ContractState} is the Ethereum smart contract persistent store, itself a mapping $\Set{Address}\rightarrow\Set{Value}$, where \Set{Address} is the set of 256-bit integers (addresses used with the \textsf{SSTORE} and \textsf{SLOAD} instructions), and \Set{Value} is the set of 256-bit values stored at and retrieved from those addresses.\footnote{We omit discussion and strict formalization of unimportant details, e.g., EOAs do not have a \Set{State} field, etc.}
  
Let $\DbState{DB}_1, \DbState{DB}_2 \in\Set{DbState}$ denote the database states for which root hashes are logged onto the blockchain ledger.
The abstract states are logged onto the blockchain using a cryptographic hash summary.
We take a canonical representation of the abstract state as a Merkle-Patricia tree, send that through a serialization---an injection function---from the tree representation to a byte stream, and send that output to the cryptographic hash function.
Since we assume that the cryptographic hash function $h(\cdot)$ is collision resistant, we will abuse notation slightly and sometimes use the hash of the serialization of the Merkle-Patricia tree representation of \DbState{DB} and the state \DbState{DB} interchangeably.
When we need to explicitly refer to the hash value, it is denoted $h(\DbState{DB})$.

Transactions are mappings from state to state.
While smart contract transactions take arguments, for our purposes here we will model them as curried functions, so when a user submits a transaction proposal with calldata, we view the calldata and the contract code as together specifying an unary state-to-state function.
All transactions change state, and we denote the type $\Set{Txn}\colon \Set{DbState}\rightarrow\Set{DbState}$.
Transactions that revert will change only the \Set{Nonce} and \Set{Balance} members of the sender account value to sequence transactions and pay for gas fees incurred, while all other \Set{DbState} members---especially smart contract accounts' \Set{ContractState} to which the ACID transaction semantics apply---are unchanged.

A sequence of transactions $\TxSeq{T}=(t_1, t_2, \ldots, t_k)$ recorded in a block moves the system from one state to the other, viz, the transactions $t_i \in \Set{Txn}$ changes account balances and smart contract persistent states, and $\TxSeq{T}$ is the composition of these mappings:
\begin{align}
  \DbState{DB}_2 &= \TxSeq{T}(\DbState{DB}_1) \\
                 &= (t_1; t_2; \ldots; t_k)(\DbState{DB}_1) \\
                 &= t_k(\ldots t_2(t_1(\DbState{DB}_1)))
\end{align}

While all transactions are logged---via the state root hash---into the append-only log, not all (intermediate) states are logged.
A blockchain block contains various payload data, including the list of transaction being executed/confirmed by the block, but only the state root hash of the state that results from the execution of \emph{all} those transactions, starting from the previous block's state, is included in the log entry.
In other words, $\DbState{DB}_2$ is in the block containing $\TxSeq{T}$, but there is no entry for $t_1(\DbState{DB}_1)$ (unless $k=1$).

Let $B_j$ denote the $j^{\text{th}}$ block in the blockchain, and let  $\TxSeq{T}_j$ denote the sequence of transaction recorded in it.
Let $\DbState{DB}_0$ be the genesis block.
Thus block $B_j$ takes the system from state $\DbState{DB}_{j-1}$ to state $\DbState{DB}_j$.

\subsection{The Block \emph{Tree}}

In full generality, we think of blocks as containing pointers to other blocks in a content-addressable storage (CAS); that is, a ``pointer'' is the cryptographic hash of the other block's contents.
These pointers uniquely refer to contents unless hash collisions can be obtained, and such data objects must form acyclic graphs---cycles are impossible unless hash preimages can be found.
In the context of blockchains, a particular block is distinguished as the genesis block, which contains no hash-based pointers, and all other blocks contain only a single hash-based parent block back pointer.
This means that in the CAS data structures view, we have a tree of blocks where blocks have parent pointers, and the genesis block is the root of the tree, since it has no parent.
We are not concerned with other trees with different root blocks---they are associated with other systems, e.g., test networks, or a separate blockchain instance created using the same software/design.

Each fork in the block tree is essentially another possible world resulting from the decision to include the block's transactions.
The block tree is the multiverse view, and the global consensus of a blockchain is the ``legitimate'' universe.

Note that in our model, two or more nodes can contain the same sequence of transactions \TxSeq{T}, but have different resultant states.
While the bogus nodes can be readily identified by validating the transactions, light clients are unable to do so since they do not maintain \Set{ContractState}.

We adopt the usual tree notation: $\depth(n)$ to denote the path length from a node $n$ to the root, $\distance(n_1,n_2)$ to denote the number of edges that must be traversed to reach $n_2$ from $n_1$, and $\isAncestorOf(n_1,n_2)$ to denote the predicate for $n_1$ being an ancestor of $n_2$.
The value of $\depth(n)$ is just the block number for $n$ if $n$ is one of the consensus blockchain blocks.
All consensus blockchain blocks $n$ will have $\isAncestorOf(\DbState{DB}_0,n)$ true.
Light clients can easily compute these functions.

\subsubsection{Blockchains are Long Skinny Block Trees}

We never talk about block trees but instead only blockchains.
Occasionally there are mentions of ``uncle'' blocks in PoW chains which indicates the potential tree structure.
And occasionally sidechains that adversaries might construct to try to fool victims.

The reason for this is due to the incentive design in blockchain systems---both PoW and PoS designs---motivate entities who can create new nodes to only grow the ``legitimate'' chain.
In PoW systems, this is the leaf with highest depth or path length, which is again a \emph{global} property.
Block rewards associated with minting a block are part of consensus reality only when the block is in the longest chain, so hashing to grown a shorter chain would just be wasting resources (and an opportunity cost).

In PoS system, only the consensus committee can create new blocks using cryptographic signatures, and bad behaviors such as equivocation that could lead to branch creation are (typically) punished by slashing.
Stake is the resource limit for the adversary: only entities with a significant amount of stake delegated to them can participate in the consensus protocol.

In the context of LRAs, \ExtractorLetter is free to add (mint) blocks arbitrarily as well as hide nodes---typically entire subtrees---in order to control what \VictimLetter sees and which possible world \VictimLetter would believe to be the consensus reality.
Of course, \ExtractorLetter is unlikely to want or need to create a block tree with many branches.
The security assumptions regarding recent validation committees says that \ExtractorLetter \emph{cannot} mint branches too close to the current actual consensus block, but ideally we want to avoid making stronger assumptions if we can.
As it turns out, this is not needed.

\subsubsection{Making Side Chains / Branches Infeasible}

Most of the approaches to LRAs attempt to make it infeasible for adversaries to create side chains that would be accepted.
This means that even if the intended victim is eclipsed and has no idea what the current blockchain state---for PoW the identity of the longest branch, and for PoS the current validator committee membership---the victim will still not be fooled.

Section~\ref{sec:RelatedWork} below goes into more detail about these approaches.

\subsection{Short \Goober Approach}

Rather than trying to make it possible to recognize that a side chain is invalid or to somehow make it impossible to create side chains, the short \goober idea takes a different approach.
Instead, we limit the potential for damage that could result if an eclipsed victim is fooled to act based on side chain information.
One key advantage is that some implementation strategies (see Section~\ref{sec:ImplementationApproaches} below) require no changes in the cryptographic primitives used nor changes in how the core blockchain messages formats or their interpretation.

A \emph{short \goober} is a mechanism to tie a transaction proposal to a node in the block tree.
This block, which is our \emph{anchor} block, contains the state information that the user relied upon to decide to make the transaction proposal.\footnote{Short \goobers are not useful if users relied on no on-chain state, since this means that the users do intend that their transaction be applicable at every potentially valid state.}
That is, the \goober asserts that the transaction is null and void unless the current block is in the causal future of the anchor block, and further, that anchor is not too far in the past of the current block, as specified by the \emph{length} or \emph{range} of the \goober.

How would this work?

More formally, short \goobers uses a higher-order function $$\gooberfunc \in \left(\Txn, \Set{BlockNumber}, \Set{Blockhash}, \UintEvm\right) \rightarrow \Txn$$ which returns a ``range limited'' version of its input transaction.
In block tree notation,
\begin{equation*}
    \footnotesize
    \gooberfunc(t, i, v, l) \mapsto
    \begin{array}[t]{l}
      \lambda s: \texttt{if ($\depth(\parent)$} < i \\
      \hspace*{1.25cm}\vee~ \texttt{$\depth(\parent)$} \ge i + l \\
      \hspace*{1.25cm}\vee~ h(\up(\parent,\depth(\parent)-i) \ne v \texttt{) \{}\\
      \hspace*{1.0cm}\texttt{revert}(s) \\
      \hspace*{0.5cm}\texttt{\} else \{} \\
      \hspace*{1.0cm}t(s) \\
      \hspace*{0.5cm}\texttt{\}}
    \end{array}
\end{equation*}
Here, \parent refers to the parent tree node of the block being constructed, $\up(n,k)$ is a function that returns the node $k$ hops up the tree from node $n$ (via parent node pointers), and \texttt{revert} is a transaction that always reverts.
NB: A leash length of zero will always revert, since the system cannot be constructing a block for which the hash value is known/specified; arithmetic modulo $\UintEvm$ is assumed to never wraparound or to abort the transaction if it occurs.

In a normal setting where \Victim is not eclipsed and decides to run $\gooberfunc(t,i,v,l)$ based on information at block $B_i$, the cost is only slightly more than running $t$ by itself.
In this case, \VictimLetter set $i$ to the block number of the current block, so when the transaction is later processed \texttt{block.number} will be slightly larger than $i$ due to delays from \VictimTPP tools for examining the current state and from transaction submission / execution delays (e.g., network delays, time $\gooberfunc(t,i,v,l)$ spent in the mempool, etc).
\VictimTP chooses $l$  to account for what \VictimTP would find an acceptable latency.
When $\gooberfunc(t,i,v,l)$ is part of a sequence of transactions \TxSeq{T} processed against an input block $\DbState{DB}_j$ where $i \le j < i+l$, $\gooberfunc(t,i,v,l) \approx t$.
Here by ``$\approx$'' we mean that the output state of the leashed transaction is the same as the output of the unleashed transaction when restricted to \Set{ContractState}, so the (minor) differences in \Set{Balance} due to gas usage is ignored.

Let us examine what happens if, on the other hand, \Victim is eclipsed and \Extractor has control of what \VictimTP sees.

Suppose \ExtractorLetter has provided \VictimLetter with a side chain---any nodes from the block tree \emph{not} on the consensus path---containing bogus transactions.
We assume that \VictimTP will simply use the bogus ``current'' consensus state---a block provided by \ExtractorLetter---and use a block explorer to examine the state of relevant contracts in making \VictimTPP decision to propose a transaction.
\VictimTPCap does \emph{not} have to verify the transactions in that block nor that the hash pointer to the parent node is valid.
\VictimTPCap \emph{must}, however, verify the Merkle proof for the examined contract state against the state root in the block.

Here, \VictimLetter's $i$ and $v$ parameters to \gooberfunc will be bogus and inconsistent with that of the real global consensus chain, and \VictimTPP submission of $\gooberfunc(t,i,v,l)$ will be essentially a no-op when processed by the real blockchain, costing her a little gas fee for reverting but incurring no real damage.
Similarly, if \ExtractorLetter had convinced \VictimTPO that the current chain state is shorter than actual so that \VictimTP will make a decision to submit $t$ based on stale information, \VictimTPP $i$ and $l$ parameters will prevent $\gooberfunc(t,i,v,l)$ from having any non-trivial effect.
In either of these cases, $\gooberfunc(t,i,v,l) \approx \texttt{revert}$.

Note that in order for \VictimLetter to do the checks that we require, \VictimTP needs to have access to a trusted computing base (TCB) for the Merkle proof verification.
If \VictimTP uses a block explorer under \ExtractorLetter's control and does no verification on \VictimTPP own of what the block explorer shows \VictimTPO, then \goobers cannot work: \ExtractorTP can provide bogus contract state values but a valid state root hash $v$ from a real consensus block---if \VictimTPP Merkle proof validation is skipped, \VictimTP will be fooled into issuing a leashed transaction $\gooberfunc(t,i,v,l)$ with $i$, $v$ parameters that are valid against the actual consensus blockchain.

\subsection{Dealing With Hard Forks}
\label{sec:DealingWithHardForks}

An important caveat for \VictimLetter's use of short \goobers is that \VictimTP will be unaware of hard forks that occurred when \VictimTP was asleep.
Being eclipsed, \VictimTP could easily be disallowed from learning this.

One might imagine that \ExtractorLetter can take advantage of hard forks that the blockchain community governance decided on taking, by eclipsing this event from \Victim.
Here, \ExtractorTP would continue extending the old chain when everyone else has moved on; \ExtractorTP is able to do so because \ExtractorTP has the validator keys from the LRA.
The converse scenario is also potentially a problem.
Here, \ExtractorLetter acquired enough cryptographic key material to falsify governance decisions for a bogus hard fork.
\ExtractorTPCap would create a bogus design / software update to the blockchain with a back door, and then falsify any on-chain governance logs to approve the update.
When \VictimLetter's light client chain validation see the this, \VictimTP is unlikely to be able to distinguish this from a real governance change.

Fortunately, a minor extension to \goobers is feasible: include the identity of the hard fork (version number or name for the chain, updated genesis block information, etc) along with the rest of the \goober parameters in the signature payload, and ensure that the extended \goober parameters encoding scheme does not change regardless of what else might change in hard forks.
In either scenario, as long as the fork version identity is included, any new transactions that \VictimTP signs will be inapplicable to the real blockchain.

Critically, this requires that \VictimLetter is able to check that this property of the signature format is preserved.
This should be feasible by including a tool that parses the blockchain/hard fork identity from generated signatures; \VictimTP can use the old software to validate the new.
Note that if \ExtractorLetter fools \VictimLetter into downloading a ``new'' (trojan) version of blockchain client software to run, \VictimTPP TCB could become compromised, and there are \emph{many} other issues that need to be addressed.
Ideally, blockchain user client software will run in a sandboxed environment~\cite{jangda2019not,10.1145/3498688,10.1145/3474123.3486762,10.1145/3429885.3429966}, so that any hard fork updated software runs in a separate sandbox from the signing tool (which should never need to be updated\footnote{This is not entirely trivial, since we have to assume that no change in signature algorithms will be needed and that the signature format etc will remain unchanged.}), so that arbitrary oracle access to the signature key or its exfiltration will be difficult.


\subsection{Interaction with Checkpoint / Replay}

Blockchain systems have bugs like any other complex software systems.
A post-deployment system management / sustaining security engineering issue is how to handle situations where bugs have been exploited, e.g., used in a zero-day attack.
In blockchain systems, making such changes require an upgrade / hard fork.
Of course, not all hard forks make drastic/dramatic changes to how a blockchain operates.
Some may only make relatively minor changes, for example, fixing a bug by changing how a particular virtual machine instruction handles an edge case.

One way to undo the damage done by zero-day exploits is to use checkpointed transactions and replay them after the zero-day vulnerabilities have been fixed~\cite{DBLP:journals/corr/abs-2201-07920}.
This potentially interacts poorly with \goobers in that the replay transaction execution should yield different states---that's the point of fixing the bug!---and thus the state root hash in the affected blocks in the blockchain will change, and furthermore propagating through children blocks because the hash-based parent pointers will also change.

From the point of view of user intentions, a \goober-protected transaction that is \emph{not} part of the zero-day attack should execute against the new bug-fixed chain state with the same expected semantics as before.
This is a policy question: a goal of blockchain smart contracts is to have deterministic execution from well defined semantics, and the users who proposed transactions affected by the bug fix probably expect that, insofar as possible, post-fork their transactions will have the same effect on the system state as they had before the bug fix.

This situation is no different, however, from how checkpoint/replay has to deal with any smart contract code that uses the \blockhash EVM instruction and is not particular to \goobers.
The hash value returned \emph{must} be the same as were returned in the earlier version of the system in order for the smart contract transaction execution to be the same.
To provide consistent deterministic execution in the face of bug-fixed replays from checkpoints, the cryptographic hash function needs to be modified.
Fortunately, designing an alternate cryptographic hash function with the same collision resistance properties while returning the desired values is relatively straightforward.

Suppose a zero-day bug was found and fixed, and community governance decides that transactions in blocks $B_j$, $j \ge z$ should be replayed, i.e., the zero day vulnerability was exploited by an adversary starting with transactions in $B_z$.
Suppose there is a total of $n$ blocks.
Since the blockchain virtual machine has changed, transactions in block $B_z$ and after will be interpreted differently.
The semantic meaning of the contract code will be (slightly) different state transformations, and the result will effectively a new/forked blockchain consisting of blocks $B_j'$ where $B_j' = B_j$ when $j < z$, but $B_j' \ne B_j$ for $z \le j \le n$.
Blocks $B_j'$ and $B_j$ differ only in the state root hash, i.e., they will be contain the same transactions in the same order, etc.
As we will see, because of the way we define a new hash function $h'$ for the forked chain, the hash pointer to the parent node in the block tree will be the same value as that from using $h$.

We define $h'(\cdot)$ for the forked blockchain using an ``output swizzler'' function $g$.
Let
\[
    g(x) \triangleq \left\{
        \begin{array}{ll}
          h(B_j) & \mbox{if $z \le j \le n$ and $x = h(B_j')$,}\\
          h(B_j') & \mbox{if $z \le j \le n$ and $x = h(B_j)$,}\\
          x & \mbox{otherwise.}
        \end{array}
    \right.
\]
then $h'(x) \triangleq g(h(x))$.
We refer to the input values for which $g(x) \ne x$ as ``swizzled inputs.''
Note that $g$ is a permutation on the space of hash output values, so it is a bijection.

NB: the values $\left\{B_0, \ldots, B_n, B_z', B_{z+1}', \ldots B_n'\right\}$ are distinct and all corresponding hash values are similarly distinct, since otherwise we would already have one or more cryptographic hash collisions, violating our assumption that $h$ is collision resistant.

\begin{theorem}
    $h'$ is collision resistant iff $h$ is collision resistant.
\end{theorem}
\begin{proof}
(Sketch.)
If we had an effective means to find distinct $x$ and $y$ such that $h(x)=h(y)$, then $g(h(x))$ and $g(h(y))$ are obviously the same value.
Conversely, suppose we had an effective means to find distinct $x$ and $y$ such that $h'(x)=h'(y)$.
This means $g(h(x))=g(h(y))$.
But since $g$ is a bijection, $h(x)=h(y)$ must hold.

The resources needed to implement an algorithm for $g$ is quite modest, since the number of swizzled inputs is linear on the number of affected blocks ($n-z$).
\end{proof}\nocite{cryptoeprint:2006:281}

Should there be another zero-day bug fix and a new fork created, the definition for that replacement hash function $h''$ is straightforward and natural.
The composition of two permutations is itself a permutation, so the proof extends naturally.

\subsection{Implementation Approaches}
\label{sec:ImplementationApproaches}

There are many ways to implement \goobers.
While \goobers are straightforward in theory, from an engineering perspective---especially when retrofitting security mechanisms---exactly how they are implemented affects usability and can be critically important.
Obviously, one could just require all top-level transaction entry points in all contracts to add the extra restriction parameters, but that would be hugely disruptive.
Similarly, requiring users to create new, ``throwaway'' contracts with fixed leash parameters and then calling them would work, but would needlessly pollute the blockchain state with throwaway contracts that will never be used again.

Below, we examine a few ways that \goobers could be realized in a less disruptive way and discuss the trade-offs.
In all cases, some degree of tooling support is likely required.
In order for users to use \goobers, they have to specify the $i$, $v$, and $l$ parameters for the top-level transaction, and this means that the top-level invocation will be different.
Using \goobers properly is impossible without either user-visible changes (new interfaces that require users to supply additional parameters) or tooling changes (tools to automatically figure out the appropriate values of $i$, $v$, and $l$).

For brevity that we omit the check for hard fork identities discussed in Section~\ref{sec:DealingWithHardForks}; the changes needed for fork-resilient \goobers are straightforward.
Making a blockchain fork identity available for verification can be implemented in a variety of ways.
The simplest is to encode the fork identity in the result returned by the \chainid EVM instruction~\cite{EIP1344}, though if there are backward compatibility issues with existing contract usage a special smart contract that returns the appropriate constant---which is updated on hard forks---could be used instead.

\subsubsection{VM Changes}

The EVM \blockhash opcode allows access to the most recent 256 blocks which may not be enough.
In order to allow larger values of $l$, the \goober length to the anchor block, we could (1) simply modify \blockhash to work even with block numbers more than 256 from the current one, or (2) introduce a new instruction, say, \blockhashex, which has the extended, ``full-domain'' semantics.

The former choice can break existing code that rely on the \blockhash op code returning zero for a block that is committed more than 256 blocks ago, and without a thorough checking of all contracts it is difficult to know whether this is safe.
The latter choice is safe, but uses a new opcode when one might not be actually needed.

Exactly how many blocks from the current blocknumber should \blockhashex work is another important design parameter.
If it is too small, then some users may find using \goobers difficult.
More importantly, a sudden influx of transactions can push transactions into later blocks, so a DOS attack could nullify \goobered transactions.

\subsubsection{Wrapper Contracts}

Rather than changing tooling or the underlying EVM semantics, for many contracts it would be reasonable to wrap the user-invoked entry points with a \goobered interface.
Contract entry points that are intended to be called by other contracts as subtransactions would not need this treatment, since the \goobering is only needed for the top-level transaction.

\begin{figure}[thpb]
    \centering
    {\small
\begin{verbatim}
// SPDX-License-Identifier: 0BSD
pragma solidity ^0.8.11;

contract Leashed {
  modifier Leash(uint256 bn,
                 uint256 hash_val,
                 uint256 leash_len) {
    require(bn < uint256(block.number),
            "used bogus future state!?!");
    require(uint256(block.number)
            < bn + leash_len + 1,
            "used stale old state!");
    require(uint256(blockhash(bn))
            == hash_val,
            "on side chain!");
    _;
  }
}
\end{verbatim}
    }
    \caption{\Goobering via a Solidity modifier}
    \label{fig:Leashing}
\end{figure}

Figure~\ref{fig:Leashing} shows a \goober implementation via a modifier, and Figure~\ref{fig:LeashedErc20} shows the \goobered interface for an ERC20 contract.
Note that the \texttt{LeashedErc20} contract is \emph{not} ERC20 compatible.

\begin{figure}[thpb]
    \centering
    {\small
\begin{verbatim}
// SPDX-License-Identifier: 0BSD
pragma solidity ^0.8.11;
import "@openzeppelin/IERC20.sol";
import "./leashed.sol";

contract LeashedERC20 is Leashed {
  IERC20 public erc_20_contract;

  constructor (address _erc20) {
    erc_20_contract = IERC20(_erc20);
  }

  function lsh_transfer(uint256 bn,
                        uint256 hash_val,
                        uint256 leash_len,
                        address to,
                        uint256 amount)
    public Leash(bn,
                 hash_val,
                 leash_len) returns(bool) {
    return erc_20_contract.transfer(to, amount);
  }
  // ... etc
}
\end{verbatim}
    }
    \caption{\Goobered Version of an ERC20 Contract}
    \label{fig:LeashedErc20}
\end{figure}

Note that this approach is ineffective beyond the \blockhash instruction's 256 block limit.
For longer \goober lengths, some other change to allow more than an hour of submission delay may still be needed, since Ethereum has a block interval of approximately 14 seconds.

\subsubsection{A Contract For An Extended \blockhash}

A non-VM change is to install a \blockhash caching contract that can provide the blockhashes for all block numbers.
While an VM change to install this as a precompiled contract (akin to other precompiled contracts like crypto support~\cite{EIP152, EIP196, EIP197}, etc) to ensure that all blockhashes are available would be most robust, it is not necessary since this contract would not need access to anything other than ordinary EVM opcodes in its implementation.
An external ``driver'' to call this contract to lookup and cache any missing blockhashes, at least once every 256 blocks, suffices, though this dependency could, in principle, turn into a denial-of-service opportunity for an adversary.
However, if many users use this for LRA prevention (and pay for it storing hash values), an external driver might be superfluous.

Note that it is easy for an adversary with a super-threshold of validator keys to falsify blockhash information in the caching contract in their side chain.
This is not a problem, however, since \Victim doesn't care if \VictimTPP transaction proposal could be made to have an effect on a side chain:  it is only the nullification at the consensus path that matters, and as noted above, as long as the Merkle proof validation for on-chain data that contributed to the decision to sign the contract proposal is done, \VictimTP is fine.

\subsubsection{A General \Goober Gateway Contract}

The approach of using the \texttt{Leash} Solidity modifier to make leashed versions of contract interfaces on a per-contract basis does not scale well.
Instead, a better approach would be a contract that could work with \emph{any} existing smart contract interfaces.

This is possible using lower-level EVM instructions.
Here the calldata would contain a prefix consisting of the leash parameters and the leash target contract address, followed by the normal calldata for the target contract call.
The generic \goober contract will verify the leash parameters, and if things are correct, extract the target contract address into an \texttt{address} variable \texttt{\_addr}, and use \texttt{\_addr.call} to pass the real calldata through.
Appendix~\ref{sec:GenericEVMShortGooberGateway} shows one implementation.

There are some drawbacks to this approach.
This could have problems with target contracts that need to use the maximum calldata length as determined by the per block gas limit.

More practically, standard wallets and tools can be changed to automatically add the leash prefixes and indirect through the generic \goober gateway.
This would essentially leave the user experience unchanged.

\subsubsection{Transaction Proposal Metadata}

Instead of making all contracts include \goober parameters as call arguments and implementing checks as explicit code or indirecting through a gateway contract, the \goober parameters could be included as part of the transaction proposal metadata.
In this design variation, the \goober parameters is included in the signature, like gas fees or gas limits, but not sent into any contracts.
Instead, before the smart contract execution the \goober parameters are validated first, and the call automatically reverted if the check fails.
Because \goobering is only needed for the top-level transaction, no change to \texttt{call}, \texttt{callcode}, or \texttt{delegatecall} to include the \goober parameters is needed.
Because this is a declarative style design, the transaction can be sequenced---that is needed to increment the EOA nonce and to charge some minimal gas fee---but no smart contract code needs to be loaded for its execution if/when the \goober parameters would have caused a revert.

Since such an implementation affects the transaction signature format, how validators / compute nodes process the proposed transaction, etc, this is not a simple change for existing smart contract blockchain networks, even though \goober parameters can be (initially) optional and phased in.
Taking this approach would require a network upgrade that introduces new functionality.

On the other hand, this approach allows existing contracts to work with short \goobers, with no changes needed those contracts.

\subsubsection{Non-Ethereum Compatible Blockchains}

\Goobers can be implemented natively in blockchain systems that do not attempt to maintain (some level of) Ethereum compatibility or even use a completely different virtual machine than the EVM.

In particular, tools to invoke transactions can provide UI to specify the \goober restriction parameters, the transaction signature scheme can include these parameters as non-call parameter metadata, etc.
If adding \goobers to the non-Ethereum blockchain is a retrofit, the same approach as with transaction proposal metadata above is still feasible: it could be done in an incremental fashion in a sequence of network updates, with a new transaction proposal format and submission endpoints etc coexisting with the old format while tooling catches up.
Removing compatibility with existing Ethereum tools as a design goal makes \goober implementation much easier, at the engineering cost trade-off of needing more custom tooling work.

\subsection{Extensions and Generalizations}

In one sense, \goobers are nothing more than ensuring that a cryptographically signed messages include the right security context information.
Ensuring signature schemes are securely used, beyond that of the security of the primitive itself, is old and includes signature padding~\cite{coron2002optimal}, ensuring message serialization is an injection, etc.
What is perhaps more important here is deciding what security context information is important, ignoring what is irrelevant, and coming up with a simple, understandable design that is easy to explain to end users.

In another sense, \goobers are nothing more than a version of correctness preconditions from Hoare logic or precondition checks commonly recommended for writing defense-in-depth code---extended to a ``desirability'' predicate or determining a $P$ in the Hoare triple $\left\{ P \right\} C \left\{ Q \right\}$ where $C$ is our transaction $t$ and $Q = \textit{"profit"}$, and the including a check for $P$.
Here, \Victim, the prototypical victim, has an off-line, private decision procedure for when \VictimTP wants to submit a transaction proposal, based on the current consensus state.
\Extractor does not know this predicate exactly, but has an approximation for it that allows \ExtractorTPO to construct an attractive side-chain universe that will cause \VictimTPO to propose a transaction.
In this view, the post-condition achieved by the transaction, when applied to the actual consensus state, is undesirable to \Victim, but desirable by \Extractor.
Obviously, \Victim's predicate has off-chain inputs that \Extractor cannot influence, e.g., \VictimTPP cash reserves.

If we understood how on-chain inputs to the off-line, private desirability predicate for executing transactions behave, a weakest precondition checking version of \goobers could be devised, where the on-chain inputs are checked before allowing the main body of the transaction to be executed.
Such a weakest precondition \goober would be less general, highly customized for users, and much harder to implement.
Additionally, such a \goober would obviously reveal information about the decision making process and may sometimes be undesirable even though such disclosures can lead to greater economic efficiencies.
In contrast, the block/hash/length constrained \goober is much easier to understand and does not require public accessors for other contracts' state, however, and the simplicity of the correctness argument makes it easy to (manually) apply universally.
An area to explore is whether automated tools for extracting precondition checks as users decide on their transactions (e.g., part of block explorer) and providing smart contract tools/libraries to allow highly flexible precondition verification could be feasible and usable.
%
%
%

\section{Related Work}
\label{sec:RelatedWork}

Other approaches to defending against LRAs~\cite{ButerinLongRangeAdaptiveProofOfWork,8525396,Deirmentzoglou2019ASO} include securing the key material via forward secure signatures~\cite{kuznetsov2021asynchronous}, additional consensus-like on-chain voting on checkpoints coupled with limits on token transfers~\cite{azouvi2020winkle}, or using another ``helper'' blockchain to supplement/enhance the security of a PoS blockchain~\cite{https://doi.org/10.48550/arxiv.2201.07946,steinhoff2021bms}.
We discuss these approaches below.

\subsection{Forward Secure Signatures}

Forward secure cryptography schemes~\cite{bellare1999forward,buchmann2011xmss} are effective in theory, but can be difficult to realize in practice: copies of a running process’s memory are made during virtual machine migration as well as during normal paging activity;\footnote{Locking pages via \texttt{mlock(2)} prevents paging, but not VM migration/failover.} filesystem data are copied as part of normal system backups; and even erasing or overwriting data is problematic due to magnetic disk head alignment~\cite{260635} or semiconductor data remanence and FLASH wear leveling~\cite{gutmann2001data,10.1007/11545262_25,Breeuwsma_forensicdata,FUKAMI2017S1}.
System level data leakages, like CPU side channels, are difficult to eradicate.

Of course, the signing functionality does not have to be realized on common or standard system configurations.
We could build a fault-tolerant service from a set of TEEs from distinct fault domains to ensure no hard-to-erase copies are made.
However, using such custom solutions would incur significant operational costs.\footnote{Both Intel SGX and AMD SEV-SNP realizations of the TEE abstraction allow paging of TEE memory, but the content is encrypted and thus should not leak information.}

More importantly, while altruistic validators might be willing to build and use an expensive configuration such as an network of TEEs, in the BAR model the profit maximizing strategy for a rational actor is to save copies of the key material to sell in an auction after they exit the ecosystem!
The security assumption needed for the forward secure approach to succeed will require that the number of non-altruistic validators in every validation committee to be sub-threshold.

\subsection{Winkle: Voting on Checkpoints and Token Transfer Limits}

The focus of Winkle is on the security of Validator Based Consensus with Reconfiguration (VBCR) systems.
The focus is on protecting the changes in the validator set as the VBCR evolves, though the approach also apply to the transactions being sequenced by the validators.

Winkle protects PoS blockchains against LRAs using two main mechanisms: on-chain checkpoint voting, and a token transfer limit to prevent an adversary from acquiring a super-threshold number of tokens in a fork.
The key idea is that a new block validates the previous block, not only because the validators do so, but because the users who propose new transactions do it as well---via the checkpoint votes.
A threshold number of checkpoint votes is needed for it to succeed.
Winkle specifies $q$ as the threshold needed as a fraction of all native tokens.
This latter form of validation makes the block a ``confirmed checkpoint.''

The first mechanism, on-chain voting, works roughly as follows.
Transactions get an extra field, used to hold a checkpoint block hash value, which is the vote being cast by the transaction proposer.
In addition to the validators reaching consensus via stake-weighted voting on new blocks containing transactions, every transaction within the block would include a vote on a previous checkpoint state.
A vote on a checkpoint also implicitly votes for all previous checkpoints in its timeline (blocks on the path to the root of the block tree).
Since users may not be active, they can also delegate their checkpoint votes to ``pools'' in roughly the same way that users can delegate their tokens to validators for staking and earn rewards.

The second mechanism, token transfer limitations, is intended to support the first mechanism.
By limiting the total amount of token transfers, Winkle ensures that an adversary attempting to create a side chain fork cannot accumulate too many tokens and thus acquire a super-threshold amount of checkpoint voting power.
The concern is that without this, the adversary can pack token transferring transactions into the first block of the side chain so that on subsequent block(s) they will have enough checkpoint voting power to confirm it.

Winkle is unclear on what should occur if the majority of the votes somehow disagreed with the blocks that the validators signed.
Winkle's formalism does not include state roots in the blocks, so in their model validators only provide transaction order finality and not state value finality~\cite{DBLP:journals/corr/abs-2201-07920}, so any disagreement would presumably be due to scheduling / front-running issues.
Without a design specification for how disagreements should be handled, perhaps ``witnessing'' better describes the role of checkpoint hashes in transactions.

\subsubsection{Long Range Assumptions}

A key difference between the threat model used in Winkle and that used in this paper is that Winkle must assume that the majority of users will keep their signature keys secure even in the long run.
This is a much stronger assumption that used above; in contrast, we assume the worst case: all keys except those belonging to the potential victim and to recent validation committee members are assumed to be compromised.

Clearly, if the users/pools have a high turnover rate or they practiced hygienic key rotation and transferred their tokens to a new accounts/signature keys periodically, then it is quite plausible that older keys will not be perpetually kept secure and can be acquired in an LRA.
While it is difficult to know whether, many years from now, what percentage of transaction signing keys will be compromisable, this potentially invalidates the assumption that checkpoint votes cannot be falsified.

In order to fully understand the security of checkpoint confirmation style designs for LRAs, there needs to be a way to relate the adversarial cost to the number of validator or user/pool keys that might be exposed.
Winkle appears to allow all validator keys to be exposed, but all or the majority of voter/pool keys may have to remain secure for all time.

\subsubsection{Incentive Compatibility}

Winkle provides incentives for honest behavior and tries to prevent concentrations of power or centralization at large pools.

The network rewards those whose vote contributed to a checkpoint being confirmed.
The reward amount is weighted and capped, so that pools are unlikely to grow too large: after a pool has reached a threshold size, delegating more votes to the pool will no longer increase the reward that it receives.
Less active users rationally seeking to maximize rewards will tend to migrate from a super-threshold pool to sub-threshold pools.

\paragraph{Sybil Pools}

The goal of discouraging the formation of large pools is to ensure there are enough independent pools for decentralization.
Unfortunately, in the BAR model, rational pools can mount Sybil attacks.

Winkle models the operational cost to pool operator $i$ as $E_i$ but does not specify how $E_i$ and $E_j$ might relate for operators $i$ and $j$.
If pools were independent, then one might imagine that $E_i$ and $E_j$ are approximately the same: they have to run validators / full clients, pay for networking, monitoring, etc.
When pools Sybil, a single real pool operator incurs the non-recurring cost to set up a full client, etc, but every subsequent Sybil pool operator can piggy back on the results, yielding a low recurring cost for setting up each additional Sybil pool operator identity.

\paragraph{Lazy Lemmings}

Even if pool operators do not mount Sybil attacks to earn rewards linear on the amount delegated, a purely rational actor (in the BAR model) may conclude that, since votes are public, rather than actually operating a validating full client and incur the cost of replicated contract execution, it will be more profitable to simply vote the way that the others are voting.
Since there is no \emph{penalty} for voting incorrectly, this can be a profit maximizing strategy---and even possibly safe for the network when there are plenty of altruistic participants for ``lazy lemming'' voters to follow.
However, if a sub-threshold number of Byzantine participants were to cast their votes quickly \emph{and} the altruistic participants votes/transactions are delayed or lost, the (short-term rational) lemming voter could be mislead about the wisdom of the crowd and help push the vote count over the threshold/cliff.

What this means is that the security assumption required for checkpoint voting to work is standard Byzantine Fault Tolerance, with $q$ fraction of votes delegated to altruistic (and not just rational) players.

While the lazy lemmings strategy is potentially a problem with all delegated PoS systems, slashing for incorrect votes can serve as a disincentive for using the follow-the-crowd strategy when only a few votes have been cast.
Other common techniques for preventing lazy lemmings include commit-and-reveal, though that would not be practical here.

\paragraph{Token Transfer Restrictions}

Winkle imposes a token transfer limit for each block (Lemma 5.3).
The purpose of this limit is to ensure that an adversary with control of two past validator set (due to LRA) cannot manipulate transactions between two valid database states so that the adversary will gain more than the maximum threshold amount of stake (by processing in-transfers but not out-transfers).
Being able to create a fork where the adversary has greater than maximum threshold means that the adversary can falsify state so that all future validator set membership will be under their control.

It is unclear how this limit is proposed to be enforced and how enforcement is to be incentivized: the obvious choices are (1) transactions that would push the per-block token transfer over the limit could be deferred until the next block(s) or (1) the transactions could be aborted.
Since the amount of transfer is not knowable, in general, until the transacton executes, it would be unfair to validators to execute a transaction (in part) until the transaction amount is known to defer it, since the gas fees will not be paid; similarly, it would be unfair to transaction proposers if their transactions were aborted---with gas fee paid---through no fault of theirs, due to global conditions that are hard to predict.

This requirement is also a global property, which makes ``trustless finality'' style estimates difficult since transfers could be either deferred or aborted for non-local reasons, and knowing whether (or when) one’s transaction will actually be committed is difficult, making the use of transfers as payments problematic.

\paragraph{Decentralized Finance and Locked Tokens}

The Winkle design does not allow tokens under the control of smart contracts to participate in checkpoint voting.
This means that tokens locked up in smart contracts such as bridges, liquidity pools, etc.
Tokens can also be locked up---but nonetheless remain in effect liquid---via schemes such as liquid staking where a service stakes depositors' tokens to earn rewards while issuing proxy tokens representing the deposit amount which can be traded, while maintaining a liquidity pool to ensure redemption of the proxy token for the native token is feasible.
The amount of locked tokens could be a significant fraction of the total token supply and could cause problems with the liveness of checkpoint voting.

\paragraph{Winkle Threat Model}

An important point is that Winkle uses a different threat model than that used in this paper.
Winkle classifies accounts as active/honest, byzantine, eclipsed, and dead.
Despite reward payments for voting for checkpoints, the scheme is not incentive compatible.
There are no (hyper-)rational actors who might sell their tokens and auction off their signature keys.
Obviously, it is not only the validators who might rationally decide to sell their keys---which were used in the BFT protocol to sign blocks---but ordinary users can also behave rationally:  after selling their tokens or transferring to a new account, they can sell their old account's signature keys.
This would render the on-chain votes suspect, since the blocks on a forked side chain would contain bogus transactions that vote for bad checkpoints.

\subsection{Using Another Blockchain}

In contrast to Winkle which tries to enhance a blockchain to secure itself,
the approach take by Babylon~\cite{https://doi.org/10.48550/arxiv.2201.07946} and BMS~\cite{steinhoff2021bms} is to essentially secure a first blockchain with a second, more stable, and hopefully more secure blockchain.
While this seems reasonable, there is the obvious question of potential infinite regress:  what makes the second blockchain actually more secure?
Does it in turn have to be secured by yet another blockchain?

As noted in Section~\ref{sec:ThreatModel}, PoW blockchains are subject to LRAs as well if the eclipsed victim can be confused about the passage of time.
A plausible side chain---using \emph{less} than 51\% hashing power---can be constructed if the victim is prevented from seeing the actual longer consensus chain.
If the victim can be eclipsed for the PoS blockchain, there is no reason why they cannot also be eclipsed with respect to the PoW blockchain that is being used to secure it.
The additional adversarial work needed to block communications to the PoW chain as well as the PoS chain would seem to be virtually nil.

BMS examines the case of a PoS blockchain being made more secured using a PoW blockchain such as Ethereum to log the configuration changes, i.e., the validation committee elections.
Babylon enhances the security of a PoS blockchain by logging configuration changes to a well-known PoW blockchain, Bitcoin.
Like most blockchain designs, the security of a PoW chain depends on it being popular and actively mined, because the adversarial cost to take over a PoW blockchain is the ``51\% attack'', where the adversary needs 50\%+$\epsilon$ of the total mining power.
If the total number of (non-Byzantine) miners drops, the cost to accrue 50\%+$\epsilon$ becomes smaller, and the PoW chain's security suffers.

However, an important rationale, if not the \emph{raison d'etre}, for the PoS blockchain design approach is to provide a better environment than PoW style designs: faster smart contract execution, lower latency, less wasteful energy usage, etc.
If the PoS blockchain really is wildly successful, then one would imagine that loads would migrate from the PoW blockchain to the PoS system.
This renders a security design that tries to inherit security from the PoS system suspect, since the success of the PoS system becomes self-defeating!

One reasonable view, of course, is that the PoS system requires more ``security support'' while it is still small, so a design like these makes sense until the PoS system becomes more mature.
However, this still leaves us in a (small) quandry:  a mature PoS system still need LRA protection, and it should not try to derive support from a PoW system in its decline.


\section{Conclusion}
\label{sec:Conclusion}

We have described the novel, yet simple technique of using short \goobers approach to prevent powerful adversaries from benefiting from long-range attacks on blockchain systems.
Abandoning the attempt to distinguish one branch of the block tree from another, we instead require users to explicitly restrict the application scope of transactions they sign; this requires solving a much simpler subproblem, resulting in a solution to the LRA problem that works well in the BAR model and is relatively easy to implement.
No new cryptographic tools are needed; the standard notion of cryptographic context suffices.

The \goobers technique is orthogonal to other approaches such as trying to recognize side chains or preventing accidental private key disclosures via forward-secure signature schemes.
Multiple techniques can be applied for additional defense-in-depth: \goobers will not help users who are careless in their use, e.g., specify the block number $i$ to be close to the root rather than the leaf node in the block tree and with an excessively large length limit $l$, whereas techniques that do not require correct user behavior would work regardless.



\renewcommand{\baselinestretch}{1.0}



{\footnotesize \bibliographystyle{acm}
\bibliography{paper}}

\appendix
\appendixpage
\addappheadtotoc
\section{Winkle Security Proof Issues}
\label{sec:WinkleSecurityProofIssues}

Winkle's security proof for Lemma 5.3 has at least two errors.
Firstly, the restriction that an adversary who acquires old validator private keys can only construct bogus transaction as subsequences is wrong.
Secondly, that the amount of tokens sent or received due to the execution of a sub-sequence of the original transactions between two states has \emph{anything} to do with the amount of tokens sent or received due to the original sequence is wrong.

\subsection{Notation}

First, we revisit the notation used in Winkle and extend it slightly.

Let $\DbState{DB}_1$ and $\DbState{DB}_2$ denote database states.
A sequence of transactions $\TxSeq{T}=(t_1, t_2, \ldots, t_k)$ from the legitimate execution history moves the system from one state to the other, viz,
\begin{align}
  \DbState{DB}_2 &= \TxSeq{T}(\DbState{DB}_1) \\
                 &= (t_1; t_2; \ldots, t_k)(\DbState{DB}_1) \\
                 &= t_k(\ldots t_2(t_1(\DbState{DB}_1)))
\end{align}

Let $\TxSet{A}(\TxSeq{T})$ denote the set of all transaction sequences constructed from transactions in $\TxSeq{T}$ that would be acceptable, i.e., transactions from any EOA are in the order determined by the nonce value used in the signature.

Let $\WAmt{S}(\TxSeq{T}, \DbState{DB})$ and $\WAmt{R}(\TxSeq{T},\DbState{DB})$ denote the amount that adversary-controlled addresses sent and received, respectively, in the the transaction sequence \TxSeq{T} when executed against state \DbState{DB}.

\subsection{Sub-sequences}

While transactions signed by any single Externally Owned Account (EOA) must be processed in order, there are no restrictions on the relative order of transactions from independent accounts.

If \TxSeq{T} consists of $k$ transactions from a single account, then
$$\TxSet{A}(\TxSeq{T}) = \left\{(), (t_1), (t_1, t_2), \ldots, (t_1, t_2,...t_{k-1}), \TxSeq{T}\right\}$$
since that account's nonce determines transaction ordering.
Clearly, $\left|\TxSet{A}(\TxSeq{T})\right| = k+1$.
(Of course, this assumes that the empty block is considered valid; many systems have additional limitations and miners/schedulers may not be able to create blocks with a small number of transactions.)

If, on the other hand, \TxSeq{T} consists of transactions from $k$ independent EOAs, then there are many different transaction schedules that could be constructed from \TxSeq{T}, of varying lengths.
Let $\TxSet{A}_j(\TxSeq{T})$ denote the subset of acceptable transaction sequences from $\TxSet{A}(\TxSeq{T})$ of length $j$.
Then
\begin{align*}
  \left|\TxSet{A}(\TxSeq{T})\right| &= \left|\TxSet{A}_0(\TxSeq{T})\right| + \ldots + \left|\TxSet{A}_{k-1}(\TxSeq{T})\right|  + \left|\TxSet{A}_k(\TxSeq{T})\right| \\
  &= 1 + k + \ldots + \frac{k!}{2} + k! + k! \\
  &= \sum_{i=0}^k\frac{k!}{i!} \\
  &< k! \sum_{i=0}^\infty \frac{1}{i!} \\
  &= k!e
\end{align*}

Transaction ordering is only required when transactions originate from the same EOA, there are many ways for an adversary to construct acceptable fake block(s) from transaction proposals from real block(s).
The adversary is not restricted to subsequences of the original order.

\subsection{Transfer Limits}

Most PoS blockchains currently operational or being designed provide the ability to run smart contracts, Turing compute programs (modulo gas limits) that can do much more than transferring tokens.
In particular, a smart contract can programmatically control the amount being transferred.
Figure~\ref{fig:Contract} is one such smart contract.

\begin{figure}[thpb]
    \centering
    {\small
\begin{verbatim}
// SPDX-License-Identifier: 0BSD
//
// Counterexample to Lemma 5.3 of _Winkle:
// Foiling Long-Range Attacks in Proof-of-
// Stake Systems_
// https://eprint.iacr.org/2019/1440.pdf
//
// Key properties: Owner can initiate
// transfer to a recipient (e.g., a
// validator); actual amount transferred
// depends on contract state (contract
// balance).  Anyone can affect contract
// state (by sending tokens to this
// contract via the default function).

pragma solidity ^0.8.11;

contract Lemma5_3 {
  address public owner;

  constructor () {
    owner = msg.sender;
  }

  function send(address payable to,
                uint amount) public {
    require(msg.sender == owner);
    if ((address(this).balance % 2) == 1) {
      to.transfer(1);
    } else {
      to.transfer(amount);
    }
  }

  receive () external payable {}
}
\end{verbatim}
    }    
    \caption{Example contract: Adversary Receive Amounts Are State Dependent}
    \label{fig:Contract}
\end{figure}

In this contract, the amount of tokens under the control of the contract influences the number of tokens transferred.
Suppose this contract was instantiated by Alice, so she is the owner.
Imagine that $\TxSeq{T}=(t_1, t_2)$, where $t_1$ is a transaction by Bob to transfer an odd number of token base units to the contract, and $t_2$ is a transaction by Alice to transfer 1,000,000 Eth to an address under the control of the LRA adversary (e.g., Alice is in cahoots with the LRA adversary).

Suppose in $\DbState{DB}_1$, the contract's balance is even and greater than 1,000,000 Eth.
This means that in $t_1(\DbState{DB}_1)$, the contract's balance is odd, and thus the amount transferred by $t_2$ is just one base unit.
This means that $\WAmt{R}(\TxSeq{T},\DbState{DB}_1) = \text{1 Eth}$.

Now, since Alice and Bob have independent EOAs, $\TxSeq{T}^* = (t_2) \in \TxSet{A}(\TxSeq{T})$ since her transaction can go through without his being executed first.
Executing $\TxSeq{T}^*$ to get $\DbState{DB}_2^* = \TxSeq{T}^*(\DbState{DB}_1)$, we see that $\WAmt{R}(\TxSeq{T}^*,\DbState{DB}_1) = \text{1,000,000 Eth}$, so $\WAmt{R}(\TxSeq{T},\DbState{DB}_1) \ne \WAmt{R}(\TxSeq{T}^*,\DbState{DB}_1)$.

The issue is that the behavior of transactions on a smart-contract blockchain system can be highly state dependent, so computing estimated bounds on $\WAmt{S}$ or $\WAmt{R}$ over all possible transaction sequences $\TxSet{A}(\TxSeq{T})$ will require essentially executing each of those sequences against the input state.
Except for design choices such as requiring the transaction proposers to pre-declare the maximum amount transferred (and to which accounts), it is unlikely that shortcuts exists for Turing complete smart contract VMs.
We know from above that $k \le \left|\TxSet{A}(\TxSeq{T})\right| \le k!e$, and it seems reasonable that in practice the size will likely be on the higher end of the range rather than the lower end.
This implies that computing $\WAmt{S}$ or $\WAmt{R}$ will be, in general, prohibitively expensive.

While normal legitimate token transfers are unlikely to be constructed to make it difficult to bound $\WAmt{R}(\cdot)$ over $\TxSet{A}(\TxSeq{T})$, the existence of contracts that makes such bounds difficult to compute---even in the absence of users/pool keys being compromised---poses a problem for the practicality of the Winkle design.

\section{Generic EVM Short \Goober Gateway}
\label{sec:GenericEVMShortGooberGateway}

Figure~\ref{fig:generic} is a potential implementation of a generic gateway contract for checking \goobers.
The assumption here is that the tooling for end users will be modified so that when the EOA account that they control initiates a top-level transaction, the transaction indirects through this gateway contract.

\begin{figure}[thpb]
    {\small
\begin{verbatim}
// SPDX-License-Identifier: 0BSD
// Generic leash checker gateway
pragma solidity ^0.8.0;

contract LeashGateway {
  function lsh_check_abi(uint256 fork_id,
                         uint256 bn,
                         bytes32 hash_val,
                         uint256 leash_len,
                         address target,
                         bytes calldata rest)
    public payable
    returns (bytes memory) {
    require(fork_id == 0xdeadbeef,
            "wrong fork");
    require(bn < uint256(block.number),
            "used bogus future state!?!");
    require(uint256(block.number)
            < bn + leash_len + 1,
            "used stale old state!");
    // modify to use internal caching fn
    require(blockhash(bn) == hash_val,
            "on side chain!");
    // gas overhead
    uint256 pass_gas = gasleft() - 0x800;
    bool success;
    bytes memory reply;

    (success, reply) = target.call{
      value: msg.value, gas: pass_gas }(rest);

    require(success, "transaction reverted");
    return reply;
  }
}
\end{verbatim}
    }
    \caption{Generic EVM Short \Goober Gateway}
    \label{fig:generic}
\end{figure}

%
Note that there is a 224 byte overhead for the leash parameters on the input call data, and the success/failure code prefix for the return data incurs a 96 byte overhead.
The additional gas cost for this as well as the instructions executed for the checks is reasonably modest.
Caching for \texttt{blockhash} is not implemented and that would allow longer length leashes while increase costs slightly.


\end{document}